\documentclass[prb,aps,groupedaddress,superscriptaddress,twocolumn,longbibliography]{revtex4-2}
\usepackage{preamble}
\begin{document}
\title{From hidden order to skyrmions:\\
Quantum Hall states in an extended Hofstadter-Fermi-Hubbard model}
\author{F.~J.~Pauw\orcidlink{0009-0006-4188-8503}}
\affiliation{\ascaddress}
\affiliation{\mcqstaddress}
\author{U.~Schollw\"ock \orcidlink{0000-0002-2538-1802}}
\affiliation{\ascaddress}
\affiliation{\mcqstaddress}
\author{N.~Goldman\orcidlink{0000-0002-0757-7289}}
\affiliation{\ulbaddress}
\affiliation{\solvayaddress}
\affiliation{\lkbaddress}
\author{S.~Paeckel\orcidlink{0000-0001-8107-069X}}
\affiliation{\ascaddress}
\affiliation{\mcqstaddress}
\author{F.~A.~Palm\orcidlink{0000-0001-5774-5546}}
\affiliation{\ulbaddress}
\affiliation{\solvayaddress}
\date{\today}
\begin{abstract}
The interplay between topology and strong interactions gives rise to a variety of exotic quantum phases, including fractional quantum Hall (FQH) states and their lattice analogs -- fractional Chern insulators (FCIs).
Such topologically ordered states host fractionalized excitations and, for spinful systems, are often accompanied by ferromagnetism and skyrmions.
Here, we study a Hofstadter-Hubbard model of spinful fermions on a square lattice, extended by nearest-neighbor interactions.
Using large-scale density matrix renormalization group (DMRG) simulations, we demonstrate the emergence of a spin-polarized $\nicefrac{1}{3}$-Laughlin-like FCI phase, characterized by a  quantized many-body Chern number, a finite charge gap, and hidden off-diagonal long-range order.
We further investigate the quantum Hall ferromagnet at $\nu=1$ and its skyrmionic excitations upon doping.
In particular, we find that nearest-neighbor repulsion is sufficient to stabilize both particle- and hole-skyrmions in the ground state around $\nu=1$, whereas we do not find such textures around $\nu=\nicefrac{1}{3}$.
The diagnostic toolbox presented in this work, based on local densities, correlation functions, and spin-resolved observables, is directly applicable in quantum gas microscopy experiments.
Our results open new pathways for experimental exploration of FCIs with spin textures in both ultracold atom and electronic systems.
\end{abstract}
\maketitle

\section{Introduction}
\begin{figure}
    \centering
    \includegraphics{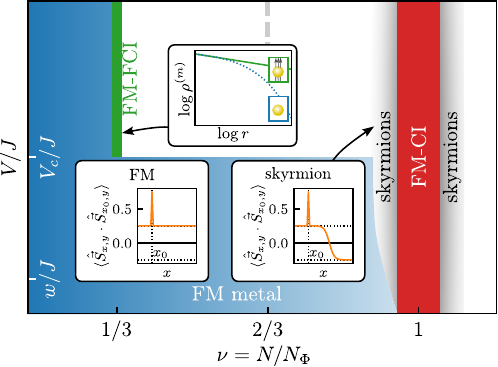}
    \caption{Parameter space studied in this work and schematic phase diagram.
    Here, $w$ is the band width of the lowest Chern band and $\nicefrac{V_c}{J}$ is the critical interaction strength needed to stabilize a fractional Chern insulator (FCI) at filling factor $\nu=\nicefrac{1}{3}$.
    ``FM'' denotes ferromagnetic phases.
    White areas indicate metallic states with short-range antiferromagnetic correlations.
    The dashed gray line indicates an incompressible unpolarized state at $\nu=\nicefrac{2}{3}$, the nature of which is not discussed in detail here.
    The insets sketch the qualitative spin-spin correlations $\braket{\hat{\vec{S}}_{x,y}\cdot\hat{\vec{S}}_{x_0,y}}$ on a cylinder and composite boson ($^m$CB) correlations $\rho^{(m)}$ on a square as defined in Eq.~\eqref{eq:hodlro} for characteristic phases.
    Note that for $x=x_0$ we expect $\braket{\hat{\vec{S}}_{x,y}\cdot\hat{\vec{S}}_{x,y}} = s(s+1) = \nicefrac{3}{4}$ based on the spin-1/2 of the constituent particles.
    \label{fig:NNInt:SketchPhaseDiagram}
    }
\end{figure}

In the last decade, lattice analogs of fractional quantum Hall (FQH) states -- so-called fractional Chern insulators (FCI) -- have emerged as a promising platform to explore topologically ordered phases of matter both theoretically~\cite{Soerensen2005,Palmer2006,Hafezi2007,Palmer2008,Moeller2009,Moeller2010,Moeller2015,Bauer2016,Huegel2017,Motruk2017,He2017,Gerster2017,Andrews2018,Dong2018,Andrews2021,Palm2021,Andrews2021a,Boesl2022,Wang2022,Palm2022} and experimentally~\cite{Clark2020,Leonard2023a,Wang2024}, even in the absence of an external magnetic field~\cite{Regnault2011,Sheng2011,Neupert2011,Wu2012}.
These phases arise in partially filled topological flat bands with interactions, giving rise to exotic quantum phenomena such as fractionally charged quasiparticle excitations with anyonic statistics.
While the details of the respective states depend on the specific lattice model, a degree of universality has been observed in their topological features~\cite{Wu2012}.
In particular, the fermionic Hofstadter-Hubbard model on a square lattice -- especially when extended with nearest-neighbor interactions -- provides a promising platform for realizing FCIs.
Here, the additional nearest-neighbor repulsion resembles the effect of the Haldane pseudo-potentials stabilizing the Laughlin-$\nicefrac{1}{3}$ state in the continuum~\cite{Haldane1983}.
Early numerical studies have demonstrated the feasibility of spin-polarized FCIs in such systems~\cite{Motruk2016}, yet a complete characterization demands robust detection strategies capable of distinguishing topological phases from competing conventional orders. 
A particularly rich aspect of FQH states is encoded in their intricate spin structure discussed theoretically~\cite{Halperin1983,Rezayi1987,Haldane1988,Lee1990,Rezayi1991,Sondhi1993,Fertig1994,Moon1995,MacDonald1996,Yang1994,Ezawa2009} and observed experimentally~\cite{Clark1989,Eisenstein1990,Engel1992,Holmes1994,Du1995,Barrett1995,Leadley1997,Leadley1998,Khandelwal1998,Kukushkin1999}.
Some quantum Hall states (e.g. at filling factors $\nu = 1$ and $\nicefrac{1}{3}$) favor ferromagnetic order even without Zeeman energy, while others (e.g. at $\nu=\nicefrac{2}{3}$) remain unpolarized or only partially polarized.
Furthermore, upon doping away from certain incompressible (F)QH ferromagnets, skyrmions may form, realizing non-trivial topological spin textures that carry both charge and spin.
Such low-energy skyrmion excitations were found in theoretical studies close to integer~\cite{Rezayi1987,Haldane1988,Lee1990,Rezayi1991,Yang1994,Sondhi1993,Fertig1994,Moon1995,MacDonald1996} and fractional filling factors~\cite{Kamilla1997,Wojs2002,Doretto2005} and experimental signatures in this direction have been observed~\cite{Barrett1995,Leadley1997,Leadley1998,Khandelwal1998}.
Their stability and efficient spin-charge coupling not only shape transport properties but also make them appealing as information carriers for spintronics~\cite{Xia2023,Yang2024}.
Understanding the interplay between interactions, topology, and spin is, therefore, crucial both from a conceptual point of view and in light of potential applications.
In this context, quantum simulator platforms provide unprecedented access to such strongly correlated topological phases in a highly tunable setting.
Quantum gas microscopy offers spatially resolved charge and spin measurements, allowing experimental access to both local correlations and global topological invariants~\cite{Bakr2009,Sherson2010,Preiss2015,Bergschneider2018,Mazurenko2017,Gross2021}.
In parallel, transition metal dichalcogenides (TMDs), exhibiting flat topological bands with valley-contrasting Berry curvature and strong interactions, have recently emerged as a compelling solid state arena for investigating FQH physics offering tunable access to FCIs including direct transport signatures and electric-field tuning~\cite{Wu2018,Wu2019,Morales2022,Cai2023,Zeng2023,Park2023}.
In this context, a possible implementation of an $\mathrm{SU}(2)$-symmetric model in an external magnetic field has been proposed~\cite{Kuhlenkamp2024}.
Together, these complementary platforms significantly expand the experimental landscape for exploring the interplay of interactions, topology, and spin.
This motivates the development of a framework that can unambiguously identify competing emergent phases.
In this work, we address the realization and identification of a variety of quantum Hall states and their accompanying spin structure in the extended Hofstadter-Fermi-Hubbard model.
To this end, we gather a toolbox of existing diagnostics which are accessible in state-of-the-art quantum gas microscopes and sufficient to unravel the coexistence of topological order and spin texture.
Using large-scale density-matrix renormalization group (DMRG) simulations, we explore the energy and spin landscape, and reveal hidden off-diagonal long-range order (HODLRO) in the composite boson basis~\cite{Girvin1987,Girvin1988,Rezayi1988,Read1989} that was recently applied to a bosonic lattice system~\cite{Pauw2024} for the first time.
At fractional filling factor $\nu=\nicefrac{1}{3}$, we confirm the presence of a lattice analog of a spin-polarized Laughlin state~\cite{Laughlin1983} of spinful fermions for sufficiently strong nearest-neighbor interaction, and find signs of an unpolarized, incompressible state at $\nu=\nicefrac{2}{3}$.
Around integer filling $\nu=1$, we present conclusive evidence of skyrmionic spin excitations, showing that nearest-neighbor repulsion is sufficient to overcome the particle-hole asymmetry found earlier~\cite{Palm2023,Ding2024}, and to stabilize hole skyrmions.
In contrast, around $\nu=\nicefrac{1}{3}$ we find no indication of skyrmion formation in the ground state irrespective of the interaction strength considered here.
The schematic phase diagram shown in Fig.~\ref{fig:NNInt:SketchPhaseDiagram} summarizes our main findings and highlights the signature observables used to probe the distinct phases.
This article is structured as follows.
We begin by introducing the model studied throughout this paper and its non-interacting properties.
In Sec.~\ref{sec:Probes} we discuss in some detail a self-contained toolbox of complementary observables both accessible experimentally and numerically, to probe the various phases of matter.
Our numerical results for both a square lattice geometry with open boundary conditions and thin cylinders are discussed in Sec.~\ref{sec:OBC} and Sec.~\ref{sec:PBC}, respectively.
We close with concluding remarks and an outlook in Sec.~\ref{sec:Conclusion}.
\section{Model and Methods}
We study spinful fermions in a Hofstadter-Hubbard model on a square lattice, extended by nearest-neighbor repulsion.
The Hamiltonian of the model in the Landau gauge reads
%
% \begin{widetext}
\begin{equation}
\begin{aligned}
    \hcH = &\sum_{\sigma}\hcH_{{\rm Hofstadter,\sigma}}\\
    &+ U\sum_{x,y} \hn_{x,y;\uparrow} \hn_{x,y;\downarrow}\\
    &+ V \sum_{x,y} \left( \hn_{x+1,y} \hn_{x,y} + \hn_{x,y+1} \hn_{x,y}\right),
    \label{eq:NNInt:Hamiltonian}
\end{aligned}
\end{equation}
% \end{widetext}
%
where $\hcH_{\rm Hofstadter, \sigma}$ denotes the single-particle Hofstadter Hamiltonian,
\begin{equation}
\begin{aligned}
    &\hcH_{\rm Hofstadter; \sigma} \\
    &= -J \sum_{x,y} \left(\hcd_{x+1,y;\sigma}\hc_{x,y;\sigma}+ \de^{2\pi\di \alpha x} \hcd_{x,y+1;\sigma}\hc_{x,y;\sigma} + \mathrm{H.c.}\right).
\end{aligned}
\end{equation}
%
% \begin{align}
%     \hcH = \sum_{\sigma}\hcH_{{\rm H,\sigma}} &+ U\sum_{x,y} \hn_{x,y;\uparrow} \hn_{x,y;\downarrow} \\ &+ V \sum_{x,y} \left( \hn_{x+1,y} \hn_{x,y} + \hn_{x,y+1} \hn_{x,y}\right) \nonumber
%     %
%     \label{eq:NNInt:Hamiltonian}
% \end{align}
%
% with
% %
% \begin{align*}
%     \hcH_{{\rm H}, \sigma} = -J \sum_{x,y} \left(\hcd_{x+1,y;\sigma}\hc_{x,y\sigma} + \de^{\di A(x)} \hcd_{x,y+1;\sigma}\hc_{x,y;\sigma} + \mathrm{H.c.}\right)
% \end{align*}
%
Here, $\hat{c}^{(\dagger)}_{x,y;\sigma}$ annihilates (creates) a spin-$\sigma$ fermion at site $(x,y)$, and $\hn_{x,y} = \hn_{x,y;\uparrow}+\hn_{x,y;\downarrow}$ is the total number operator of particles.
%
% Here $A(x)=2\pi\alpha x$ denotes the Peierls phase.
%
We fix the flux per plaquette to $2\pi\alpha = \nicefrac{2\pi}{4}$ in all our simulations.
Furthermore, we fix the Hubbard repulsion to $\nicefrac{U}{J}=8$ for the remainder of this work.
Such interactions were found to be sufficient to stabilize a ferromagnetic CI in earlier studies~\cite{Palm2023,Ding2024}.
Throughout this work, we use natural units where \mbox{$\hbar = c =1$} and consider constituent particles of charge $q=1$.

To explore the low-energy physics of the interacting model, we employ the strictly single-site density matrix renormalization group (DMRG3S) method~\cite{White1992,Schollwock2011,Hubig2015} implemented in the \textsc{SyTen}-toolkit~\cite{Syten}.
We exploit the $\mathrm{U}(1)_n$-symmetry associated with particle number conservation in all our simulations.
Whenever possible, we also enforce the $\mathrm{U}(1)_{z}$-symmetry associated with the spin perpendicular to the lattice plane, focusing on the sectors with $S_z = 0$ and $\nicefrac{N}{2}$ with $N$ the number of fermions.
By restricting to the $S_z=0$ sector, we can access all $\mathrm{SU}(2)$ multiplets and thus determine the total spin of the ground state.
By contrast, the $S_z=\nicefrac{N}{2}$ sector provides clearer signatures in the charge sector of the spin-polarized system.
In our simulations, we use bond dimensions up to $\chi=8192$ to reach convergence of our observables.

We analyze the low-energy physics across four representative cuts in the parameter space ($\nicefrac{V}{J},\nu$), where \mbox{$\nu\equiv \nicefrac{N}{N_{\Phi}}$} denotes the magnetic filling fraction relative to the lowest Chern band with $N_{\Phi}$ the number of flux quanta piercing the system.
Our simulations are performed on both a square box with hard walls \mbox{($L_x=L_y=10$)} and a cylinder \mbox{($L_x=24,L_y=4$)} with periodic boundary conditions along the $y$-direction.
Before discussing the interacting model in detail below, we briefly comment on the energy scales of the non-interacting model, see also Appendix~\ref{app:SingleParticle}.
For the cylinder, we find a lowest bulk band of width $w_{\rm cylinder}\approx 0.1 J$ consisting of $22$ states with Chern number $C=1$.
On the square lattice with open boundaries, we can identify a bulk band of width $w_{\rm square} \approx 0.3J$ hosting 20 states, despite the single-particle spectrum being less clean.
Due to the comparatively long edges along all directions, there is no sizable gap to the edge states.
The band width $w$ is the relevant energy scale when choosing interaction strengths below.

\section{Detecting \& probing\\(fractional) Chern insulators}
\label{sec:Probes}
%
% \todo[color=green,inline]{Re-explain motivation for toolbox: stimulate exp. progress by providing a comprehensive set of (realistically accessible, existing) observables\\
% Also, explain that everything is experimentally relevant, esp. energetic landscape as numerical proxy for (exp. acc.) incompressibility}
%
To probe and characterize the various ground state phases of Eq.~\eqref{eq:NNInt:Hamiltonian}, we put together a toolbox of complementary observables which are readily implemented with existing experimental techniques.
Some of these observables probe the local features, whereas others rely on global properties of the state.
Combining these enables an unambiguous identification of (F)CI states and their associated spin textures in future experiments.
\subsection{Energetic landscape}
A simple yet insightful characteristic of the system under study is its energetic landscape.
Here, we first focus on the ground state energy $E_N$ as a function of the particle number -- and hence the filling factor $\nu$.
Based on this, we can determine the charge gap
\begin{equation}
    \Delta\mu(N) = E_{N+1} + E_{N-1} - 2E_{N},
\end{equation}
which provides a numerical proxy for the experimentally accessible ground state incompressibility.
This diagnostic is particularly meaningful on very thin cylinders, where finite-size effects gap out the edge modes and the low-energy excitations reflect true bulk properties.
In contrast, on the square lattice geometries the emerging gapless edge modes obscure the bulk incompressibility, and $\Delta\mu(N)$ may vanish even for topological ordered phases.
The degeneracy originating from the gapless edge modes could in principle be lifted by applying an external potential, as is routinely done in quantum gas experiments~\cite{Binanti2024,Ünal2025}.

To avoid artificially favoring a specific $\mathrm{SU}(2)$-spin sector, we focus on the sector with \mbox{$S_{z} = 0$ ($0.5$)} for an even (odd) number of particles, respectively.
This allows us to extract the total spin of the ground state via
\begin{equation}
    \begin{aligned}
        S(S+1) &= \sum_{\mathbf{i},\mathbf{j}} \braket{\hat{\vec{S}}_{\mathbf{i}} \cdot \hat{\vec{S}}_{\mathbf{j}}}\\
        &= \sum_{\mathbf{i},\mathbf{j}} \braket{\hat{S}_{\mathbf{i}}^z \hat{S}_{\mathbf{j}}^z + \frac{1}{2} \left(\hat{S}_{\mathbf{i}}^{+} \hat{S}_{\mathbf{j}}^{-} + \hat{S}_{\mathbf{i}}^{-} \hat{S}_{\mathbf{j}}^{+}\right)}.
    \end{aligned}
\end{equation}
While the energy is mainly of interest to numerical studies, the total spin can also be extracted using quantum gas microscopes.
%
% \todo[color=lightgray,inline]{I think it should be sufficient to look at $\braket{S_i^zS_j^z}$ and assume $\mathrm{SU}(2)$ invariance. Might be relevant for experiments as well!}
%

%
Using the total spin, we can define the spin polarization $\mathcal{P}=\nicefrac{S}{S_{\rm max}}$ with $S_{\rm max}=\nicefrac{N}{2}$.
Assuming that adding holes (or particles) to a ferromagnetic state at $\nu_0$ causes $\mathcal{S}$ \mbox{(or $\mathcal{A}$)} spin flips in the system, we can parametrize the spin polarization as~\cite{Barrett1995,Khandelwal1998}
\begin{equation}
    \label{eq:spinPolarization}
    \mathcal{P} = 1 + 2 \begin{cases}
        \left(\frac{1}{\nu} - \frac{1}{\nu_0^{+}}\right) \mathcal{S} & \text{for } \nu > \nu_0^{+}\\
        0 & \text{for } \nu_0^{-} \leq \nu \leq \nu_0^{+}\\
        - \left(\frac{1}{\nu} - \frac{1}{\nu_0^{-}}\right) \mathcal{A} & \text{for } \nu < \nu_0^{-}
    \end{cases},
\end{equation}
where $\nu \in \left[\nu_0^-, \nu_0^+\right]$ is the spin-polarized region.

\subsection{Probing the charge sector}
While energy and spin observables provide valuable insights into the candidate state, identifying non-trivial order requires a complementary analysis of the charge sector.
To reduce computational complexity and maximize the resolution of experimentally accessible observables, we focus our simulations on the spin-polarized sector $S_z=\nicefrac{N}{2}$.
Prime examples in quantum gas microscopes are the local density $n(x,y) = \braket{\hat{n}_{x,y}}$ and its correlations.
The response of the local density to changes in the flux per plaquette, for instance, gives access to the many-body Chern number on a square lattice geometry with open boundaries via St\v{r}eda's formula~\cite{Streda1982}.
In contrast, for thin cylinders, FQH states are known to evolve into charge density wave states~\cite{Tao1983,Rezayi1994}.
Furthermore, in the FQH regime density-density correlations exhibit a characteristic drop at short distances, revealing the presence of a correlation hole.
Beyond simple density measurements, studying various correlation functions can provide significant insight.
For example, gapless metallic states display a power-law decay of fermionic correlations $\braket{\hat{c}_i^{\dagger} \hat{c}_j}$ whereas such correlations decay exponentially in the bulk of a gapped, topologically ordered phase.
Finally, combining these bare fermionic correlations with flux-attachment arguments can uncover the hidden ordering of the ground state.
\subsubsection*{St\v{r}eda response}
The Hall conductivity $\sigma_H$ -- the fundamental signature of FQH states -- is directly related to the many-body Chern number $\mathcal{C}_{\rm mb}$ via $\nicefrac{\sigma_H}{\sigma_0}=\mathcal{C}_{\rm mb}$ and for an incompressible phase can be obtained from simple bulk density measurements under varying flux employing St\v{r}eda's formula
\begin{align}\label{eq:Streda}
    \mathcal{C}_{\text{St\v{r}eda}} \equiv \frac{\partial n_{\text{Bulk}}}{\partial \alpha}=\frac{\sigma_H}{\sigma_0},
\end{align} 
where in natural units $\sigma_0 = \nicefrac{1}{(2\pi)}$~\cite{Streda1982}. 
In the bulk, we expect a St\v{r}eda response of $1$ ($\nicefrac{1}{3}$) for the (F)CI states at $\nu=1$ ($\nicefrac{1}{3}$) in the bulk.
Even though St\v{r}eda's arguments are rooted in the thermodynamic limit, they have been explicitly verified for dilute, finite lattices hosting FQH states~\cite{Repellin2020}.
The method has also been successfully applied to probe a minimal two-particle bosonic Laughlin state with ultra-cold atoms~\cite{Leonard2023a} and can be easily accessed numerically by computing the density profile of the ground state for a fixed number of particles with varying flux $\alpha$. 
\subsubsection*{Tao-Thouless states}
On sufficiently thin cylinders of circumference $L_y a$, the topologically ordered FQH states are known to evolve into symmetry breaking Tao-Thouless (TT) states~\cite{Tao1983,Rezayi1994,Seidel2005} while non-FCI-like states typically show uniform or weakly modulated density patterns.
As a result, pronounced changes in the density profile across a phase transition can help distinguish between different phases of matter.
For the $\nicefrac{1}{3}$-Laughlin state, the corresponding TT state features a charateristic occupation pattern of the lowest Landau-level orbitals: \mbox{$\ket{\mathrm{TT}_{1/3}} = \ket{\hdots 100100\hdots}$}.
In continuum systems on an infinite cylinder, these orbitals are centered at \mbox{$x_n = -\frac{2\pi\ell_B^2}{L_y a} n$} such that the TT charge density wave has a wavelength \mbox{$\lambda_{\mathrm{TT},1/3} = \frac{6\pi\ell_B^2}{L_y a}$} and wave vector \mbox{$ak_{\mathrm{TT},1/3} = \nicefrac{2\pi\alpha L_y}{3}$}.
Translating this to the lattice model with parameters $L_y=4$ and $\alpha=\nicefrac{1}{4}$ studied here, where \mbox{$\ell_B^2 = \frac{a^2}{2\pi \alpha}$}, yields
\begin{equation}
    ak_{\mathrm{TT},1/3} = \frac{2\pi}{3} \qquad \Leftrightarrow \qquad \lambda_{\mathrm{TT},1/3} = 3a.
\end{equation}
\subsubsection*{Density-density correlations}
An additional diagnostic tool accessible through the density of the ground state is the normalized equal-time density-density correlation function for two points $\mathbf{i},\mathbf{j}$ on the lattice
\begin{align}
    g^{(2)}_{\mathbf{i},\mathbf{j}}
    = \frac{\braket{:\hat{n}_{\mathbf{i}}\hat{n}_{\mathbf{j}}:}}{\braket{\hat{n}_{\mathbf{i}}} \braket{\hat{n}_{\mathbf{j}}}}
    = \frac{\left\langle \hat{n}_{\mathbf{i}}\hat{n}_{\mathbf{j}} \right\rangle}{\left\langle \hat{n}_{\mathbf{i}} \right\rangle \left\langle \hat{n}_{\mathbf{j}} \right\rangle} -\frac{\delta_{\mathbf{i},\mathbf{j}}}{\left\langle \hat{n}_{\mathbf{i}} \right\rangle},
\end{align}
which is routinely measured in quantum gas microscopes~\cite{Bakr2009,Sherson2010,Preiss2015,Bergschneider2018}.
As a function of the Euclidean distance $r=|\mathbf{i}-\mathbf{j}|$, the two-point correlator provides insight into the spatial distribution of the particles beyond their mean density.
For example, the Laughlin state exhibits a pronounced suppression of $g^{(2)}(r)$ at short distances, i.e. $g^{(2)}(r)\stackrel{r\to0}{\rightarrow} 0$, reflecting the formation of a correlation hole characteristic of incompressible quantum liquids~\cite{Kamilla1997,Girvin1999}.
Furthermore, the correlator smoothly approaches unity at larger distances and saturates, indicating the absence of long-range density order.

\subsubsection*{Hidden correlations}
In systems with open boundary conditions, we diagnose topological order through characteristic, yet hidden, multi-particle correlations.
To this end, we analyze two-point correlation functions of composite bosons (CB) of the form
\begin{align} \label{eq:hodlro}
    \rho^{(m)}_{\mathbf{i},\mathbf{j}}=\left\langle f_m\left( \{\hat{n}_{\mathbf{l}}\}_{\mathbf{l}\neq\mathbf{i},\mathbf{j}} \right) \hat{c}^{\dagger}_{\mathbf{i}}\hat{c}^{\nodagger}_{\mathbf{j}} \right\rangle,
\end{align}
where
\begin{align}
    f_m\left( \{\hat{n}_{\mathbf{l}}\}_{\mathbf{l}\neq\mathbf{i},\mathbf{j}} \right)=\prod_{\mathbf{l}\neq\mathbf{i},\mathbf{j}}\de^{\di m \arg\left(\frac{z_{\mathbf{l}}-z_{\mathbf{i}}}{z_{\mathbf{l}}-z_{\mathbf{j}}}\right)\hat{n}_{\mathbf{l}}},
\end{align}
realizes a singular gauge transformation with $z_{\mathbf{j}}=x_{\mathbf{j}}+\di y_{\mathbf{j}}$.
The $^m$CBs are effective degrees of freedom of $m$ flux quanta attached to each of the fermionic particles which then experience a reduced effective magnetic field. 
Thus, the function $f_m$ in Eq.~\eqref{eq:hodlro} reduces to the identity for $m=0$, while for $m=1$ and $m=3$, it stems from attaching one and three flux quanta, respectively, to the underlying fermions.
For an odd number of flux quanta these weakly interacting quasi-particles are bosonic and thus may condense~\cite{Girvin1987,Girvin1988,Rezayi1988,Read1989}.
The condensation of the $^m$CBs is then reflected in the emergence of \mbox{(quasi-)}long-range order in their correlations, which decay algebraically as $\propto r^{-2/m}$, where $r$ is the Euclidean distance of $\mathbf{i}$ and $\mathbf{j}$; this long-distance behavior refereed to as hidden off-diagonal long-range order (HODLRO) is independent of the gauge choice for the underlying magnetic vector potential.
Hidden correlations of this type were recently shown to be a powerful tool in tensor network studies of bosonic systems with open boundary conditions and are also experimentally accessible in quantum gas microscopes~\cite{Pauw2024}.
In principle, the emergence of HODLRO does not depend on the choice of boundary conditions, and the composite-boson construction can also be formulated on cylindrical systems. 
However, on a cylinder, evaluating the non-local phase factor  requires a careful treatment of branch cuts, and extracting HODLRO for periodic-boundary MPS therefore poses an interesting question left for future work.

The fermionic $\nicefrac{1}{3}$-Laughlin state is believed to be well-captured by composite bosons of one fermion with three flux quanta attached ($^3$CB).
The $^3$CBs experience no effective magnetic field and may form a condensate exhibiting (quasi-)long-range order with algebraically decaying correlations $\propto r^{-2/3}$ with increasing distance $r$.
Similarly, the $\nu=1$ state can be understood as a condensate of $^1$CBs, for which we also expect an algebraic decay for the $^1$CB correlations $\propto r^{-\kappa}$.
The exact value of the exponent $\kappa$, however, is less clear in this case.
While topologically ordered states are prominently characterized by the absence of long-range order in the fermionic basis~\cite{Girvin1987}, metallic correlations decay algebraically.
This behavior is robust under flux attachments, since the presence of the Fermi surface persists irrespective of the chosen representation.
The expected decay of the characteristic correlation functions across the different phases is summarized in Table~\ref{tab:CorrelationsDecay}.

\begin{table}[t]
    \centering
    \begin{tabular}{c|c|c||c|c|c}
        $\nu$ & $V$ & phase & $m=0$ & $m=1$ & $m=3$ \\\hline
        \multirow{2}{*}{$\nicefrac{1}{3}$} & $V = 0$ & metal & alg.~decay & alg.~decay & alg.~decay \\
        & $V > V_c$ & FCI & $\sim \mathrm{e}^{-x^2/\xi^2}$ & $\sim \mathrm{e}^{-x^2/\xi^2}$ & $\sim r^{-2/3}$ \\\hline
        \multirow{2}{*}{$1$}    & $V = 0$ & CI & $\sim \mathrm{e}^{-x^2/\xi^2}$ & $\sim r^{-\kappa}$ & $\sim \mathrm{e}^{-x^2/\xi^2}$ \\
        & $V > 0$ & CI & $\sim \mathrm{e}^{-x^2/\xi^2}$ & $\sim r^{-\kappa}$ & \text{$\sim \mathrm{e}^{-x^2/\xi^2}$} \\
    \end{tabular}
    \caption{
        Expected decay of (fermion ($m=0$) and composite boson ($m=1,~ 3$)) correlation functions for different filling factors $\nu$ and interaction strength $V$.
        Here, $V_c$ is the critical interaction strength needed to stabilize the $\nicefrac{1}{3}$-Laughlin state at $\nu=\nicefrac{1}{3}$ and $m$ refers to the number of flux quanta attached in the composite boson picture.
        $\xi$ is the correlation length and $\kappa$ the exponent of the algebraic decay for $^1$CBs at $\nu=1$.
    }
    \label{tab:CorrelationsDecay}
\end{table}

\subsection{Probing skyrmionic spin textures}
\begin{figure}[b!]
    \centering
    \includegraphics{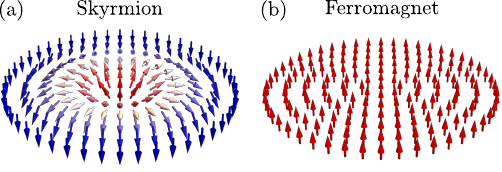}
    \caption{
        Magnetization texture $\braket{\hat{\vec{S}}_{\mathbf{i}}}$ of (a) a classical skyrmion spin texture and (b) a ferromagnet.
        In the skyrmion, local spin alignment evolves into spin reversal at larger distances.
        In contrast, in the ferromagnet the spin alignment persists at all distances.
    }
    \label{fig:SkyrmionSketch}
\end{figure}

Finally, we are interested in exotic spin textures in the vicinity of incompressible quantum Hall ferromagnets, where earlier numerical~\cite{Rezayi1987,Haldane1988,Lee1990,Rezayi1991,Yang1994,Sondhi1993,Fertig1994,Moon1995,MacDonald1996} and experimental studies~\cite{Barrett1995,Leadley1997,Leadley1998,Khandelwal1998} have found signatures of skyrmions.
These spin structures are characterized by local spin alignment and global spin anti-alignment or, more formally, a non-vanishing Pontryagin index
\begin{equation}
    Q = \frac{1}{8\pi}\int\dd^2 r~ \epsilon_{ij}~ \vec{m}(\mathbf{r}) \cdot \left(\partial_{r_i} \vec{m}(\mathbf{r}) \times \partial_{r_j}\vec{m}(\bf{r}) \right),
\end{equation}
where $\vec{m}(\mathbf{r})$ is the local magnetization~\cite{Fradkin2013}.
A typical magnetization texture of a classical skyrmion ($Q=1$) is depicted in Fig.~\ref{fig:SkyrmionSketch}(a), in contrast to a topologically trivial ($Q=0$) ferromagnetic configuration, Fig.~\ref{fig:SkyrmionSketch}(b).
In a quantum-mechanical system, a preferential direction can be chosen by measuring the spin-spin correlations relative to a reference site, so that $\braket{\hat{S}_{\mathbf{i}} \cdot \hat{S}_{\mathbf{j}}}$ gives insight into the characteristic spin texture of a skyrmion.

On thin cylinders, the corresponding spin-spin correlations are characterized by a continuous change along the cylinder axis from spin alignment to anti-alignment.
To this end, we return to the $\mathrm{U}(1)_z$-symmetric system and extract the normalized, site-resolved spin-spin correlations,
\begin{equation}
    \frac{\braket{\hat{\vec{S}}_{x,y_0} \cdot\hat{\vec{S}}_{x_0,y_0}}}{\braket{\hat{n}_{x,y_0} \hat{n}_{x_0,y_0}}},
    \label{eq:SpinSpinCorrs}
\end{equation}
relative to a reference site $(x_0, y_0)$ in the center of the system to detect skyrmions in our simulations.
Such spin-spin correlations are also routinely measured using quantum gas microscopy, for example, in the context of the doped Fermi-Hubbard model~\cite{Preiss2015,Mazurenko2017,Bergschneider2018,Gross2021}, thereby providing the unique opportunity for \textit{in-situ} observation of skyrmionic spin textures.

\section{Open boundaries: Square geometry}\label{sec:OBC}
We start our exploration of the ground-state phase diagram of the (extended) Hofstadter-Hubbard model in the square-lattice geometry with open boundaries.
In particular, we study systems of size $L_x=L_y=10$ at fixed flux per plaquette $2\pi \alpha=\nicefrac{2\pi}{4}$ and perform a scan of the parameter space sketched in Fig.~\ref{fig:NNInt:SketchPhaseDiagram}.
Here, we purely focus on the spin-polarized regime to establish the presence of the (F)CI phases.
All DMRG calculations in this section were obtained with a bond dimension of $\chi=2000$.

\subsection{Summary of main findings}
We find a robust $\nu=1$ CI phase with many-body Chern number $\mathcal{C}_{\rm mb}=1$ and quasi-long range order in the $^1$CB basis independently of the nearest-neighbor interaction strength cut.
At fractional filling $\nu=\nicefrac{1}{3}$, a phase transition from a metallic phase to a FCI phase is observed around $\nicefrac{V_c}{J}\approx 0.5$.
The FCI phase is characterized by a non-trivial Chern number $\mathcal{C}_{\rm mb}=\nicefrac{1}{3}$, strong suppression of density-density correlations at short distances and the emergence of HODLRO in the $^3$CB basis, manifesting in a power-law decay of $\rho^{(3)}(r)$.
In contrast, the metallic phase shows algebraic decay in $\rho^{(m)}(r)$ independently of the CB basis, along with enhanced short range density-density correlations.
We discuss our results in detail below.
\subsection{Results and discussion}

\subsubsection*{Probing the charge sector:\\Densities \& correlations in spin-polarized systems}
\paragraph*{St\v{r}eda response.---}
\begin{figure}[t]
    \centering
    \includegraphics{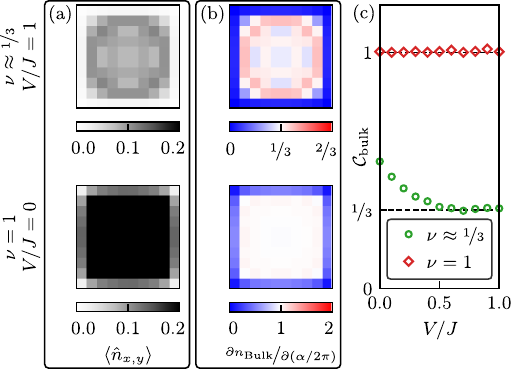}
    \caption{
        (a) Ground state density profile and (b) St\v{r}eda response for $\nu\approx \nicefrac{1}{3} \ (N=6),\ \nicefrac{V}{J}=1$  (top) and \mbox{$\nu\approx 1 \ (N=20),\ \nicefrac{V}{J}=0$} (bottom).
        In both cases, we find an extended bulk region with a many-body Chern number consistent with continuum predictions in the thermodynamic limit.
        (c) The extracted many-body Chern number for varying values of $\nicefrac{V}{J}$ shows the transition towards an FCI above \mbox{$\nicefrac{V_c}{J} \approx 0.5$}, whereas the CI persists at all interaction strengths.
        }
    \label{fig:NNInt:Streda}
\end{figure}
As a first probe of the ground state's topological nature, we examine the response of the bulk density to variations in the flux per plaquette piercing the lattice.
As discussed in Sec.~\ref{sec:Probes}, this response is directly related to the many-body Chern number through St\v{r}eda's formula,~Eq.~\eqref{eq:Streda}.

Here, we extract the Chern number for the ground states in the (F)CI parameter regime. 
To this end, we compute the mean density $\left\langle \hat{n}_{x,y} \right\rangle$ of the ground state for varying flux values $\alpha\in\{0.24,0.25,0.26\}$.
Afterwards, we perform a linear fit for all bulk latticelattice points with the Chern number $\mathcal{C}_{\rm mb}$ given by the slope.
For the incompressible (F)CI phase the density forms a droplet with a well-defined bulk plateau of density $\langle n_{\rm bulk}\rangle = \alpha\nu$.
We therefore identify the bulk region as the set of sites where the density is homogeneous, which naturally excludes the outermost sites that constitute the edge of the system.
The results are summarized in Fig.~\ref{fig:NNInt:Streda}.

For the case with nearest-neighbor repulsion \mbox{$\nicefrac{V}{J} = 1$} at \mbox{$\nu=\nicefrac{1}{3} \ (N=6)$} we find an extended bulk region with $\mathcal{C}_{\text{St\v{r}eda}}\approx \nicefrac{1}{3}$.
Averaged over this region, we find a many-body Chern number of
\begin{align}
    \mathcal{C}_{\rm bulk}=0.33\pm 0.05
\end{align}
which is in perfect agreement with the expected \mbox{$\mathcal{C}_{\rm mb}=\nicefrac{1}{3}$} for the Laughlin state in the thermodynamic limit.

Similarly, for the conjectured CI state we find an even clearer extended bulk with
\begin{align}
    \mathcal{C}_{\rm bulk}=1.00\pm0.01,
\end{align}
again in perfect agreement with the expected $\mathcal{C}_{\rm mb}=1$.

At fractional filling, varying the interaction strength $\nicefrac{V}{J}$, a phase transition from the metallic phase to the FCI phase can be observed around $\nicefrac{V_c}{J}\approx 0.5$.
At unit filling, the many-body Chern number remains stable throughout the full interaction range, suggesting the stability of the CI phase independently of the nearest-neighbor interactions.

\paragraph*{Density-density correlation.---}
\begin{figure}[t]
    \centering
    \includegraphics{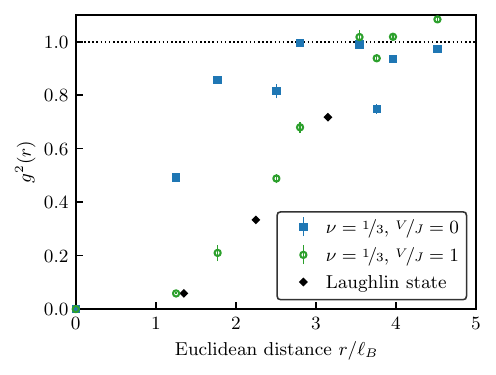}
    \caption{
        Two-particle density correlations as function of the Euclidean distance in units of the magnetic length \mbox{$\ell_B = \nicefrac{a}{\sqrt{2\pi \alpha}}$.}
        The reference site is fixed at the center of the square lattice.
        Error bars correspond to a standard deviation of the average over equidistant lattice points.
        In the case of nearest-neighbor interactions ($\nicefrac{V}{J}=1$, green circles), we observe the characteristic drop of the correlation function at short distances in qualitative agreement with the expected correlation hole for the Laughlin state (black diamonds).
        In contrast, the pure Hofstadter-Hubbard model ($\nicefrac{V}{J}=0$, blue squares) does not exhibit the characteristic polynomial drop.
        }
    \label{fig:NNInt:nncorr}
\end{figure}
\begin{figure*}[ht]
    \centering
    \includegraphics{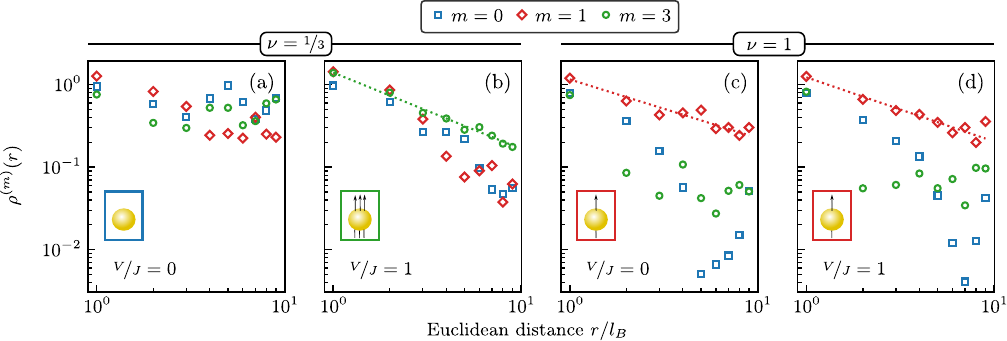}
    \caption{
        (Hidden) correlations for different filling factors and interaction strengths for the square lattice of size $L=10$ as a function of the Euclidean distance.
        The error of the sampling (standard deviation $\nicefrac{\sigma_{\rho^{(m)}(r)}}{\sqrt{N_{\mathrm{snaps}}}}$) is smaller then the marker size for all data points and thus not explicitly shown.
        The dotted lines in panels (b-d) indicate the algebraic fit of the data and the insets show the effective degrees of freedom displaying quasi long-range order in the different phases.
        While we observe the emergence of algebraically decaying correlations in the $^1$CB or $^3$CB in most cases, we find metallic correlations independent of the basis choice for $\nu\approx\nicefrac{1}{3}$ ($N=6$) at $\nicefrac{V}{J}=0$.
        These results are consistent with the expected behavior in Tab.~\ref{tab:CorrelationsDecay}.
        }
    \label{fig:NNInt:HODLRO}
\end{figure*}
Next, we turn to the analysis of density-density correlations. 
For a Laughlin-like FCI at filling $\nu = \nicefrac{1}{3}$, the correlation function is expected to exhibit several characteristic features.
First of all, due to the formation of a correlation hole, $g^{(2)}(r)$ is expected to be strongly suppressed for small $r$.
To extract $g^{(2)}(r)$, we fix the reference site at the center of the square lattice and average over all equidistant lattice points.
Our findings confirm the strong suppression and we find a short-range scaling in good agreement with the continuum expectation as shown in Fig.~\ref{fig:NNInt:nncorr}, where the continuum data was obtained via Monte Carlo sampling of the Laughlin state for $N=12$ particles on a disk.

At intermediate distances, $g^{(2)}(r)$ is expected to increase smoothly towards unity.
The weak oscillations observed are remnants of the finite lattice spacing.
At even larger distances, the density-density correlations remain at unity without developing long-range periodic order, consistent with the incompressible and liquid-like nature of the Laughlin state.
In contrast, $g^{(2)}(r)$ of the ground state of the pure Hofstadter-Hubbard model ($\nicefrac{V}{J}=0$) does not show the characteristic polynomial drop, consistent with its metallic nature.
On all scales, our finite size open boundary calculations are in good qualitative agreement with continuum expectations, supporting the conclusion that the ground state is well described by the Laughlin state.
For completeness, we show $g^{(2)}(r)$ for additional points in parameter space in App.~\ref{app:disk}.

\paragraph*{Hidden correlations.---}
To further strengthen our claim that finite interactions at fractional filling stabilize a topologically ordered lattice analog of the $\nicefrac{1}{3}$-Laughlin state, we investigate the hidden off-diagonal long-range order (HODLRO) of the ground state~\cite{Girvin1987,Girvin1988,Rezayi1988,Read1989,Pauw2024}.
To this end, we sample correlation functions in the $^m$CB basis with \mbox{$m\in\{0,1,3\}$} as described in Eq.~\eqref{eq:hodlro} from the ground state employing a \textit{perfect sampling} scheme for matrix product states~\cite{Ferris2012,Pauw2024}.
We draw an experimentally realistic number of \mbox{$N_{\rm snaps}=2\times10^3$} snapshots and focus on four characteristic points in parameter space, $\nicefrac{V}{J}=0,~ 1$ and $\nu=\nicefrac{1}{3},~ 1$.
In all cases, we fix the reference site on the edge, $\mathbf{i}=(0,L_y/2)$, and compute $\rho^{(m)}(r)$ for $m=0,1,3$, where again $r$ is the Euclidean distance.
The results are shown in Fig.~\ref{fig:NNInt:HODLRO}.
We start our discussion with the partially filled Chern band, $\nu=\nicefrac{1}{3}$.
Here, in the pure Hubbard model, the ground state is expected to be metallic, which is consistent with our numerical results showing conventional long-range correlations for $m=0$, see Fig.~\ref{fig:NNInt:HODLRO}(a).
Changing to the composite boson picture does not change the qualitative behavior, as the fermions themselves are already the effective degrees of freedom.
Switching on sufficiently strong interactions (\mbox{$\nicefrac{V}{J} = 1> \nicefrac{V_c}{J}\approx 0.5$}), we can stabilize a qualitatively different ground state exhibiting no conventional long-range order in the plain fermionic basis ($m=0$), see Fig.~\ref{fig:NNInt:HODLRO}(b).
However, we now find evidence of long-range order in the $^3$CB basis, with the expected algebraic decay of the bosonic correlations with an exponent $-\nicefrac{2}{m}$, with $m=2.20\pm{0.21}$, see the dotted fitting line in Fig.~\ref{fig:NNInt:HODLRO}(b). 
While the exponent strongly deviates from the analytical continuum prediction $m=3$, this is not surprising considering the finite-size and lattice-based nature of the system studied here~\cite{Pauw2024}.
%
% As already noted in earlier studies~\cite{Pauw2024}, this exponent is sensitive to finite-size effects 
%($\alpha=\nicefrac{1}{4}$)
% and details such as the choice of reference site.
%
Nonetheless, the observed algebraic decay provides strong evidence for quasi-long-range order in the  $^3$CB basis and supports the claim of a $\nicefrac{1}{3}$-Laughlin-like ground state.
We remark, that in the $^1$CB basis correlations still fall off exponentially, consistent with Tab.~\ref{tab:CorrelationsDecay}.
Next, we turn to the fully filled Hofstadter band, \mbox{$\nu=1$}.
Here, we find that the ground state of the pure Hubbard model exhibits strong signatures of $^1$CB condensation, in agreement with what is expected for the non-interacting CI phase, see Fig.~\ref{fig:NNInt:HODLRO}(c).
This behavior essentially carries over to the interacting case, see Fig.~\ref{fig:NNInt:HODLRO}(d), in agreement with our findings regarding the many-body Chern number and the expectation that interactions should not destroy the CI.

\section{Periodic boundaries: Cylinder geometry}\label{sec:PBC}
We complement our studies of the open boundary geometry by simulations of finite length ($L_x = 25$) cylinders with circumference $L_y = 4$ at flux $2\pi\alpha=\nicefrac{2\pi}{4}$ per plaquette.
In these systems, we especially explore the spin sector of the model, thereby revealing exotic spin-textures.
For the spin-polarized ground-state simulations presented in this section, we employed a bond dimension of $\chi=2000$, while the numerically more demanding spinful simulations required bond dimensions up to $\chi=8192$.
\subsection{Summary of main findings}
Our findings on the cylinder confirm the general picture obtained on the square lattice geometry.
In particular, we conclude that already density measurements alone are able to distinguish the substantially different phases encountered here.
Furthermore, the symmetry breaking charge density wave state realizes a thin-cylinder analog of the $\nicefrac{1}{3}$-Laughlin state.
Finally, we find evidence for the presence of skyrmion-like excitations in the particle-doped regime of the CI at $\nu=1$.
Adding nearest-neighbor repulsion restores the particle-hole symmetry of the phase diagram and leads to the formation of hole-skyrmions, an effect which can be understood as a combination of a finite spin-stiffness and interactions.
In contrast, around $\nu=\nicefrac{1}{3}$, we do not observe signs of skyrmions in the ground state on either side of the ferromagnetic insulator.
In addition to these polarized states, we also find evidence for unpolarized incompressible states at $\nu=\nicefrac{1}{2}$ and $\nicefrac{2}{3}$.
While the state at $\nu=\nicefrac{1}{2}$ is likely to be an artifact of the cylinder geometry, the latter state might be related to a spin-balanced Halperin state in the 2D-limit~\cite{Halperin1983,Pichler2025,Kuhlenkamp2025}.

\subsection{Results and discussion}
\subsubsection*{Energetic landscape:\\Ground state energy \& spin polarization}
Contrary to the open boundary geometry, we start our discussion of the cylinders by providing an overview of the energetic landscape in the $S_z=0~(0.5)$ sector for even (odd) numbers of particles~\footnote{For an additional discussion of the spin-polarized sector see Appendix~\ref{appendix:cylinder:spinpolarizedEnergetics}.}.
To stabilize our numerical results, we supplement our simulations by ground state searches in the $S_z=\nicefrac{N}{2}$ sector for those points in parameter space where we find first signs of ferromagnetic ordering.
Note that in these regimes the energy difference between the $S_z=\nicefrac{N}{2}$ and $0$ states are very small, making convergence of our simulations especially challenging in the $S_z=0$ case.

\begin{figure}
    \centering
    \includegraphics{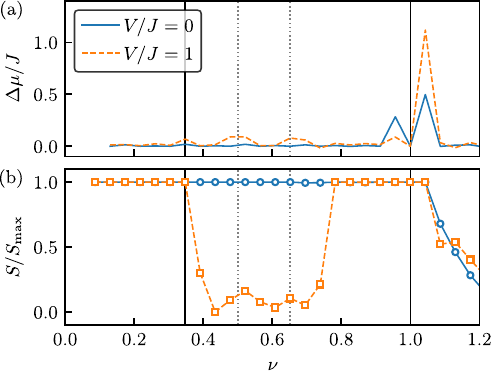}
    \caption{
        (a) Charge gap $\Delta\mu_N$ and (b) ground state spin polarization as a function of the filling factor on a cylinder.
        At $\nu=1$ we find an incompressible ground state with the charge gap on the order of the single-particle gap.
        For finite nearest-neighbor interactions $\nicefrac{V}{J}=1$ we furthermore find signals of incompressibility at $\nu\approx\nicefrac{1}{3}, \nicefrac{1}{2}$, and $\nicefrac{2}{3}$.
        %, which are absent in the non-interacting case.
        %
        % While the states at $\nu\approx\nicefrac{1}{3}$ and $\nicefrac{2}{3}$ are consistent with Laughlin states, the origin and nature of the state at $\nu\approx\nicefrac{1}{2}$ is less clear.
        %
        The pure Hubbard case exhibits ferromagnetism for all filling factors $\nu \leq 1$ and gradually depolarizes above.
        In contrast, additional nearest-neighbor interactions favor ferromagnetism below $\nu=\nicefrac{1}{3}$ and $1$, respectively, with weakly polarized states in between.
    }
    \label{fig:cylinder:EnergeticLandscape}
\end{figure}

As a first observable, we consider the charge gap $\Delta\mu(N)$ as a proxy for the incompressibility of the ground state, see Fig.~\ref{fig:cylinder:EnergeticLandscape}(a).
We find incompressible states at $\nu=1$ irrespectively of the interaction strength.
Here, the absence of a single peak at exactly $\nu=1$ is a result of the degeneracy of the edge states due to the inversion symmetry of the model, which splits the peak into the double peak structure observed here.
The signal is less pronounced in the case of pure Hubbard interactions ($\nicefrac{V}{J}=0$) than in the extended-Hubbard case, where the charge gap is roughly comparable to the band gap, $\Delta\mu(\nu=1) \approx \Delta$.
%
% The double-peak structure in $\Delta\mu(N)$ close to $\nu=1$ is a result of the existence of two low-lying edge states in the non-interacting spectrum.
%
Nevertheless, our results indicate the stability of an incompressible quantum Hall state at $\nu=1$ already for the pure Hubbard model, in agreement with earlier studies~\cite{Palm2023,Ding2024}.

Furthermore, we find that finite nearest-neighbor interactions ($\nicefrac{V}{J}=1$) stabilize incompressible states close to $\nu=\nicefrac{1}{3}$, $\nicefrac{1}{2}$ and $\nicefrac{2}{3}$ with a charge gap on the order $\approx 0.2 J$.
We note that the incompressible state at $\nu=\nicefrac{1}{2}$ is likely to be an artifact of the cylinder geometry, since the $\ket{\hdots1010\hdots}$-CDW forming in this case has a large gap upon adding an additional particle due to the nearest-neighbor repulsion.
In contrast, the unpolarized state at $\nu=\nicefrac{2}{3}$ might be related to states observed~\cite{Clark1989,Eisenstein1990,Engel1992,Du1995,Holmes1994,Kukushkin1999} and studied numerically~\cite{Xie1989,Maksym1989,Chakraborty1990a,Pichler2025} in the continuum.
We leave a more detailed characterization of the states found here to future studies and focus on the spin-polarized candidates for skyrmion formation at $\nu=1$ and $\nicefrac{1}{3}$.

Next, we consider the ground-state spin polarization $\nicefrac{S}{S_{\rm max}}$ in Fig.~\ref{fig:cylinder:EnergeticLandscape}(b).
This quantity allows us to estimate the number of spin flips upon adding or removing particles via Eq.~\eqref{eq:spinPolarization}.
A large number of flipped spins can be interpreted as a signature of skyrmion formation and was used in solid-state experiments accordingly~\cite{Barrett1995,Khandelwal1998}.
We emphasize, however, that here we do not rely exclusively on this signal to identify skyrmionic excitations, but also directly probe the local spin textures as is also possible with quantum gas microscopes.

\begin{table}[t!]
    \centering
    \begin{tabular}{c||c|c||c|c}
        \ & \multicolumn{2}{c||}{$\mathcal{A}$} & \multicolumn{2}{c}{$\mathcal{S}$} \\
        $\nicefrac{V}{J}$ & $0$ & $1$ & $0$ & $1$ \\\hline
        $\nu_0=1$ & $0.01(3)$ & $3.1(4)$ & $3.7(2)$ & $4.2(9)$ \\
        $\nu_0=\nicefrac{1}{3}$ & --- & $0.006(5)$ & --- & $0.88(9)$ \\
    \end{tabular}
    \caption{
        Number of spin flips $\mathcal{A}$ (or $\mathcal{S}$) upon adding a hole (or particle) to a ferromagnetic ground state at $\nu_0$ for a system with Hubbard interaction strength $\nicefrac{U}{J}=8$ and nearest-neighbor repulsion $V$.
        The values were obtained by fitting the model in Eq.~\eqref{eq:spinPolarization} to the spin polarization data in Fig.~\ref{fig:cylinder:EnergeticLandscape}(b).
    }
    \label{tab:cylinder:skyrmionSize}
\end{table}

For the vanilla Hubbard model ($\nicefrac{U}{J}=8,~ \nicefrac{V}{J}=0$), we find the ground state to be fully spin-polarized for $\nu\lesssim 1$.
Above $\nu=1$, the ground state gradually depolarizes in agreement with earlier studies~\cite{Palm2023,Ding2024}, hinting towards the formation of finite-size skyrmions.
The differences between $V/J=0$ and $1$ are attributed to the challenging convergence in this regime.
Using the simple model defined in Eq.~\eqref{eq:spinPolarization}, we find that each added particle induces $\mathcal{S} = 3.7(2)$ spin flips, see Table~\ref{tab:cylinder:skyrmionSize}.
In contrast, adding holes to the system does not result in any significant spin flips, $\mathcal{A} = 0.01(3)$, realizing a substantial particle-hole asymmetry.

Turning on additional nearest-neighbor repulsion, $\nicefrac{V}{J}=1$, we observe a richer spin polarization pattern between $\nu=\nicefrac{1}{3}$ and $\nu=1$.
We interpret this as the result of two different mechanisms.
Close to $\nu=1$, the ground state with nearest-neighbor repulsion is still very similar to the ground state of the Hubbard model, however the additional nearest-neighbor repulsion reduces the spin polarization also for sufficiently hole-doped systems.
We thus find ferromagnetism for $\nu\approx 1$ and potential skyrmion formation on both sides around it ($\mathcal{A}=3.1(4)$, $\mathcal{S}=4.2(9)$).
In contrast, around $\nu=\nicefrac{1}{3}$ the nearest-neighbor interactions lead to an entirely different ground state, putatively an FCI.
This state in turn favors ferromagnetic descendants for $\nu\leq \nicefrac{1}{3}$ and a finite number of spin flips as particles are added, $\mathcal{S}=0.88(9)$, whereas no such effect is visible upon hole-doping, $\mathcal{A}=0.006(5)$.

Looking ahead to the discussion of spin-spin correlations and Fig.~\ref{fig:cylinder:SpinCorrelations_V1} below, we remark that the additional nearest-neighbor interaction seems to be sufficient to stabilize both hole- and particle-skyrmions around $\nu=1$ in good agreement with experimental results in the continuum~\cite{Barrett1995,Leadley1998,Khandelwal1998}.
The effect of further-range interactions to stabilize skyrmions as lowest-energy states on the hole-doped side in continuum systems was discussed in Refs.~\cite{MacDonald1996,Wojs2002} and can heuristically be understood as a competition of the finite spin-stiffness and the interaction energy.
Around $\nu=\nicefrac{1}{3}$, where earlier theoretical studies showed that the stability of skyrmions depends sensitively on the details of the model~\cite{Kamilla1997,Wojs2002}, the situation is less clear at this stage and the fate of FQH skyrmions has to be investigated carefully.
In particular, in the composite fermion (CF) picture~\cite{Jain1989} the effective interactions of CFs at $\nu=\nicefrac{1}{3}$ are much weaker than for fermions at $\nu=1$ and might therefore be insufficient to stabilize skyrmions as lowest-energy states.
In the next sections, we will provide further evidence that the incompressible states are indeed lattice analogs of (fractional) quantum Hall states and that the excitations around $\nu=1$ feature skyrmionic spin textures, whereas at $\nu=\nicefrac{1}{3}$ the ground state does not seem to host skyrmions in the systems investigated here.
We would like to emphasize that at the other incompressible fillings $\nu=\nicefrac{1}{2}$ and $\nicefrac{2}{3}$ we do not find signs of similar spin features.

\subsubsection*{Probing the charge sector:\\Densities \& correlations in spin-polarized systems}
\paragraph*{Density.---}
To deepen our understanding of the nature of the incompressible phases observed in the energetic and spin sectors, we now turn to an analysis of the charge degrees of freedom.
We focus our analysis on the partially filled Hofstadter band at $\nu=\nicefrac{1}{3}$.
\begin{figure}[t]
    \centering
    \includegraphics{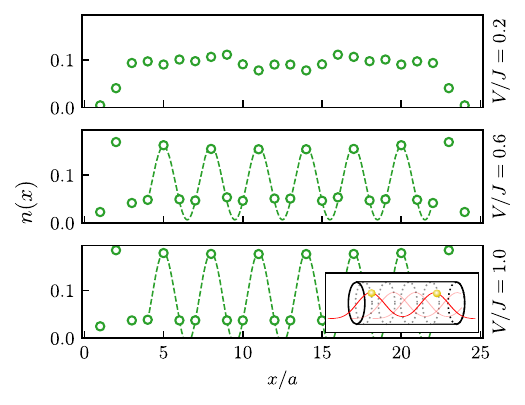}
    \caption{
            Position-dependent ground state density $n(x)$ for different values of $\nicefrac{V}{J}$ at $\nu=\nicefrac{1}{3}$.
            For sufficiently large interaction, we observe the formation of a charge density wave with the characteristic pattern of a Tao-Thouless state with wave vector $k_{\rm TT,1/3}$.
            The dashed lines are sinusoidal fits using the expected wave vector.
            The inset in the lowest panel shows the occupation of the lowest Landau level orbitals in the Tao-Thouless state.
    }
    \label{fig:NNInt:CDW}
\end{figure}
\begin{figure}[t]
    \centering
    \includegraphics{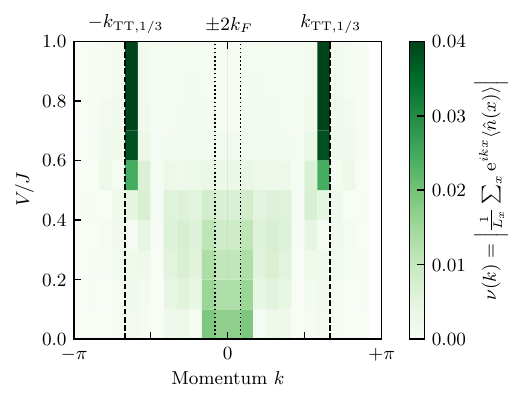}
    \caption{Fourier amplitudes of the $x$-dependent density as a function of the interaction strength $\nicefrac{V}{J}$ at $\nu=\nicefrac{1}{3}$.
    At interaction $\nicefrac{V}{J}\gtrsim0.5$ the TT wave vector $k_{\mathrm{TT},1/3}$ emerges as the dominant Fourier mode.
    }
    \label{fig:NNInt:FourierModes}
\end{figure}
In the quasi-one-dimensional cylinder geometry, we begin by evaluating the ground state density as a first diagnostic in this direction.
Due to the periodic boundary conditions along $y$, it is sufficient to consider the $x$-dependent density
\begin{equation}
    n(x) = \frac{1}{L_y} \sum_{y=1}^{L_y} \braket{\hat{n}_{x,y}}.
\end{equation}
In Fig.~\ref{fig:NNInt:CDW}, we show exemplary density profiles $n(x)$ for different values of the nearest-neighbor interaction strength \mbox{$\nicefrac{V}{J}=0.2,~0.6,~1.0$}.

In the limit of weak nearest-neighbor repulsion, \mbox{$\nicefrac{V}{J}=0.2$}, the ground state density does not exhibit any pronounced density oscillations.
In contrast, the strongly interacting system shows clear signs of a charge density wave.
Furthermore, the bulk features of the charge density wave are accurately captured by the simple fit function
\begin{equation}
    n^{\mathrm{fit}}(x) = \frac{N\alpha}{L_x} + n_1 \sin(k_{\mathrm{TT},1/3} x + \phi_0)
\end{equation}
with the CDW amplitude $n_1$ and a phase shift $\phi_0$ as the only free parameters.
We conclude that for sufficiently strong interactions the ground state is accurately described as a (lattice) Tao-Thouless state as discussed in Sec.~\ref{sec:Probes}.

To further investigate the change in behavior as the interaction strength is increased, we investigate the Fourier amplitudes
\begin{equation}
    \nu(k) = \left|\frac{1}{L_x} \sum_x \mathrm{e}^{\di kx} n(x)\right|
\end{equation}
of the density, see Fig.~\ref{fig:NNInt:FourierModes}.
In agreement with the open boundary results, we also observe a transition from the metallic to the FCI phase around $\nicefrac{V_c}{J} \approx 0.5$.
We find that for sufficiently strong interactions, $\nicefrac{V}{J}\gtrsim0.5$, the dominant Fourier mode is coming from the Tao-Thouless wave vector $k_{\mathrm{TT},1/3}$.
In contrast, in the weakly interacting regime a substantial part of the Fourier weight originates from momenta $\approx \pm 2k_F$ with $\nicefrac{2k_F}{\pi}=\nicefrac{N}{L_xL_y} = \bar{n}$ in agreement with Friedel oscillations for weakly interacting \mbox{(quasi-)one}-dimensional fermions of density $\bar{n}$~\cite{Dolfi2015}.

\begin{figure}
    \centering
    \includegraphics{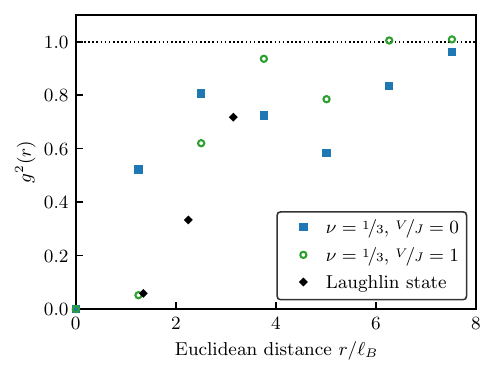}
    \caption{
        Two-particle density correlations as function of the Euclidean distance in units of the magnetic length \mbox{$\ell_B = \nicefrac{a}{\sqrt{2\pi \alpha}}$.}
        In the case of nearest-neighbor interactions ($\nicefrac{V}{J}=1$, green circles), we observe the characteristic drop of the correlation function at short distances in qualitative agreement with the expected correlation hole for the Laughlin state.
        In contrast, the pure Hofstadter-Hubbard model ($\nicefrac{V}{J}=0$, blue squares) does not exhibit the characteristic polynomial drop.
        Our findings are in agreement with continuum results for the Laughlin state (black diamonds).
    }
    \label{fig:NNInt:nn_corr_cyl}
\end{figure}

\paragraph*{Density-density correlations.---}
Analogously to the square lattice geometry, the characteristics of the density-density correlations in the thin cylinder limit offer an alternative diagnostic of the various phases.
Fig.~\ref{fig:NNInt:nn_corr_cyl} shows the normalized equal-time density-density correlation function $g^{(2)}(r)$ as a function of Euclidean distance $\nicefrac{r}{\ell_B}$ at filling $\nu = \nicefrac{1}{3}$, comparing the pure Hubbard case ($\nicefrac{V}{J} = 0$) with the extended model ($\nicefrac{V}{J} = 1$) along the x-direction for a single $y$-cut. 
The reference site is fixed at the center of the system.
In the presence of nearest-neighbor interactions ($\nicefrac{V}{J} = 1$), we observe the expected suppression of $g^{(2)}(r)$ at short distances.
The correlation function rises towards unity at larger distances, without developing long-range periodic modulations, consistent with an incompressible, liquid-like ground state.
We find that our numerical data is in good agreement with continuum results for the Laughlin state.
These observations provide strong numerical evidence that the ground state for $\nicefrac{V}{J} = 1$ realizes a lattice version of the Laughlin state, exhibiting the expected short-range repulsion and incompressible character.
The short-range suppression is absent in the pure Hubbard model.
%, which instead shows elevated short-range correlations and lacks the characteristic structure of a fractionalized phase.
%

\subsubsection*{Probing skyrmionic spin textures: Spin-spin correlations}
\begin{figure*}[t]
    \centering
    \includegraphics{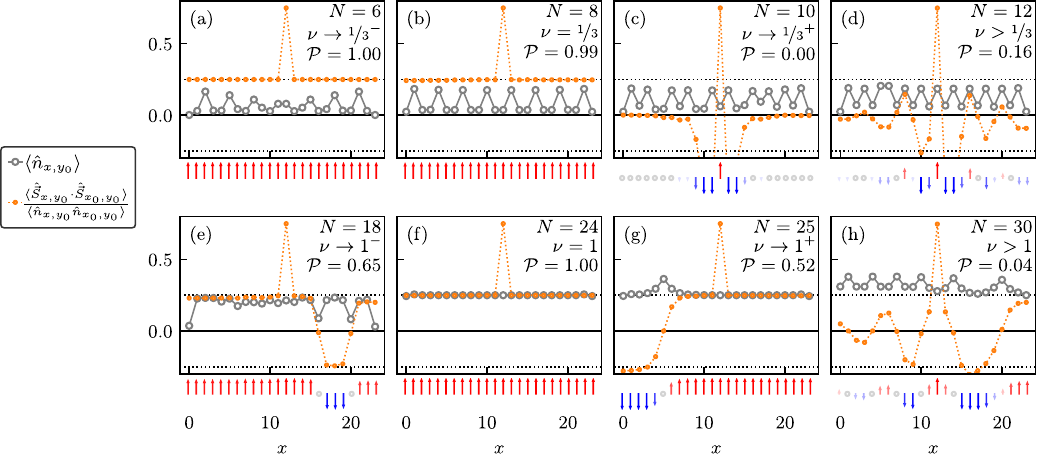}
    \caption{
        Normalized spin-spin correlations for $\nicefrac{U}{J}=8$, $\nicefrac{V}{J}=1$ and different numbers $N$ of particles.
        Below each panel, the corresponding spin configuration is sketched.
        % 
        % close to the fractional filling factor $\nu=\nicefrac{1}{3}$ (a-d) and close to the integer filling factor $\nu=1$ (e-h).
        %
        We observe fractional quantum Hall ferromagnetism for $\nu\lesssim\nicefrac{1}{3}$ (a,~b), which on the particle-doped side gives way to short-range spin anti-alignment with uncorrelated spins at larger distances (c,~d).
        %
        % Upon adding even more particles, the correlations exhibit short range spin anti-alignment (d).
        %
        For even larger numbers of particles, at $\nu=1$ (f) we find long-range ferromagnetic order, which evolves into non-trivial, skyrmion-like spin textures carried by localized excitations upon both \mbox{hole- (e)} and \mbox{particle-doping (g)}.
        Finally, substantially above $\nu=1$, the spins once again become essentially uncorrelated at larger distances with some non-trivial short-range behavior (h).
        Note that for $\mathbf{i}=\mathbf{j}$ we expect $\braket{\vec{S}_{\mathbf{i}}\cdot\vec{S}_{\mathbf{i}}} = s(s+1) = \nicefrac{3}{4}$ based on the spin-1/2 of the constituent particles.
    }
    \label{fig:cylinder:SpinCorrelations_V1}
\end{figure*}
Having established the emergence of a CI at $\nu=1$ and an FCI at $\nu=\nicefrac{1}{3}$, we now turn towards a more detailed study of their accompanying spin textures.
As we have seen before, see Fig.~\ref{fig:cylinder:EnergeticLandscape}(b), the ferromagnetic insulators feature a characteristic depolarization upon doping away from the incompressible point.
This depolarization can be further understood by means of the normalized, site-resolved spin-spin correlations in Eq.~\eqref{eq:SpinSpinCorrs} for a reference site $(x_0, y_0)$ in the center of the system.

As the ground state phase diagram is richer for the extended Hubbard model ($\nicefrac{U}{J}=8$, $\nicefrac{V}{J}=1$), we focus on this model for our analysis.
Additional data for the vanilla Hubbard model can be found in Appendix~\ref{appendix:cylinder:pureHubbard:spinSector}.
We furthermore strictly constrain our simulations to the $S_z=0$ ($0.5$) sector, thus not artificially stabilizing the ferromagnetic order around $\nu=\nicefrac{1}{3}$ and $1$.
This, in turn, allows us to characterize the spin structure of low-lying, charged excitations on top of the spin-polarized insulators.

As discussed before, we find long-range ferromagnetic order for $\nu = \nicefrac{1}{3}$ also visible from the spin-spin correlations, see Fig.~\ref{fig:cylinder:SpinCorrelations_V1}(b).
Doping away from the incompressible FCI we still find ferromagnetic order on the hole-doped side (Fig.~\ref{fig:cylinder:SpinCorrelations_V1}(a)), whereas the particle-doped side exhibits a short-ranged spin anti-alignment with uncorrelated spins at larger distances.
%
% Due to the formation of Tao-Thouless (TT) states on thin cylinders~\cite{Tao1983}, we expect quasi-holes and quasi-particles to manifest as domain walls between different TT configurations.
%
% Indeed, we observe such domain walls for $N=7$ and $9$, see Fig.~\ref{fig:cylinder:SpinCorrelations_V1}(a,~c), with $N=8$ being exactly at $\nu=\nicefrac{1}{3}$.
%
% The domain walls are accompanied by spin textures similar to the ones found for $\nu\gtrsim 1$, reminiscent of skyrmions.
%
% In particular, through such a spin texture the spin alignment is reversed with local spin alignment being restored away from the domain wall.
%
% Note that the spin-spin correlations hint towards the formation of skyrmions on both sides of $\nu=\nicefrac{1}{3}$, however the small number of particles in this regime makes definitive statements difficult, especially on the hole-doped side.
%
In particular, we do not find skyrmions in the ground state of the system, however these might still be low-lying excitations which could potentially be stabilized by a suitably chosen interaction.

Adding more particles drives the system into a regime with not as easily interpretable spin textures, see Fig.~\ref{fig:cylinder:SpinCorrelations_V1}(d).
Generally, however, the states for $\nu\approx 0.4 - 0.7$ are dominated by local spin anti-alignment, reminiscent of a Pauli exclusion hole in weakly-interacting fermionic systems.

Increasing the particle number even further, we recover long-range ferromagnetic order for $\nu \approx 1$, see Fig.~\ref{fig:cylinder:SpinCorrelations_V1}(f).
In contrast, for filling factors sufficiently far away from $\nu=1$ we observe the formation of hole- (Fig.~\ref{fig:cylinder:SpinCorrelations_V1}(e)) and particle-like excitations (Fig.~\ref{fig:cylinder:SpinCorrelations_V1}(g)), respectively.
These excitations are visible in the excess charge around the $\bar{n}=\alpha$ background and carry non-trivial spin textures reminiscent of skyrmions.
In particular, through such a spin texture the spin alignment is reversed with local spin alignment being restored away from the domain wall.
We emphasize that hole skyrmions -- absent in the pure Hubbard model -- are stabilized by the additional nearest-neighbor interaction.
In contrast, particle skyrmions also emerge in the pure Hubbard model, see Appendix~\ref{appendix:cylinder:pureHubbard:spinSector}, as already observed earlier~\cite{Palm2023,Ding2024}.

Upon particle-doping the system further away from the QH insulator ($\nu \gg 1$), the spin correlations show short-ranged spin anti-alignment with spins uncorrelated at larger distances, see Fig.~\ref{fig:cylinder:SpinCorrelations_V1}(h).
This behavior is again reminiscent of a non-interacting spinful fermions exhibiting a Pauli correlation hole.
We note, however, that in this regime convergence of the simulations is generally difficult to achieve, so that we cannot exclude the emergence of more sophisticated structures here.

We conclude by remarking that we expect qualitatively the same behavior to emerge in systems with open boundaries, where the correlations studied here are directly visible in spin-resolved quantum gas microscopy.
To resolve this phenomenology, however, would require to gap out the edge modes, for example by applying an external potential.

\section{Conclusion and Outlook}
\label{sec:Conclusion}
In this work, we investigated the spinful fermionic Hofstadter-Hubbard model on a square lattice with additional nearest-neighbor interactions with both open and cylindrical boundary conditions.
Employing a variety of experimentally accessible diagnostic tools -- ranging from local observables to topological markers -- we studied qualitatively the phase diagram shown in Fig.~\ref{fig:NNInt:SketchPhaseDiagram}.
At filling factor $\nu = \nicefrac{1}{3}$, we observed a robust transition to a lattice analog of the Laughlin-$\nicefrac{1}{3}$ state at sufficiently strong nearest-neighbor interactions ($\nicefrac{V}{J}\gtrsim 0.5$).
This FCI phase is characterized by a non-trivial many-body Chern number, incompressibility, and hidden off-diagonal long-range order in the composite boson basis.
While we do not find skyrmionic excitations around the $\nu=\nicefrac{1}{3}$ FCI, we provide evidence that the additional nearest-neighbor repulsion is sufficient to stabilize hole-skyrmions at $\nu=1$, which were elusive in earlier studies~\cite{Palm2023,Ding2024}.
Furthermore, we confirm the existence of particle-skyrmions in the ground state for $\nu\gtrsim1$.
These topological spin excitations arise as a consequence of doping the ferromagnetic insulator, leading to characteristic spin-spin correlations.
Our results highlight the rich confluence of  topology, interactions and spin and pave the way towards experimental explorations of skyrmionic excitations in quantum simulation platforms such as quantum gas microscopes and TMDs.
Studying the fate of FQH skyrmions around $\nu=\nicefrac{1}{3}$ at even stronger interactions poses an interesting subject for follow-up studies, as does a more detailed investigation of the unpolarized states at $\nicefrac{1}{2}$ and $\nicefrac{2}{3}$.
In the context of the potential formation of a Halperin state at $\nicefrac{2}{3}$, a spin-resolved variant of the hidden-order paradigm might prove useful.

Additionally, future work could examine the spatial extent of individual skyrmions in more detail, as well as the interaction energy between them as a function of distance -- key steps towards understanding their collective behavior and potential for topological spin transport.
On the numerical side, extending the toolbox by incorporating probes such as flux insertion techniques or entanglement-based diagnostics -- such as resolving orbital, spatial, or particle entanglement spectra~\cite{Li2008,Sterdyniak2011,Sterdyniak2012} -- could provide deeper insight into the topological order, edge physics, and internal structure of excitations in FCI phases.
In particular, resolving the effect of spinful excitations on the entanglement spectrum poses an interesting challenge at the intersection of quantum many-body physics and quantum information theory.
\begin{acknowledgments}
    \ 
    The authors would like to thank F.~Grusdt, J.~L\'eonard, R.~Rosa-Medina, C.~Van~Bastelaere, L.~Vanderstraeten, and K.~Viehbahn for fruitful discussions.
   FJP, US and SP acknowledge support by the Deutsche Forschungsgemeinschaft (DFG, German Research Foun- dation) under Germany’s Excellence Strategy-426 EXC- 2111-390814868.
    Work in Brussels is supported by the ERC Grant LATIS, the EOS project CHEQS, and the Fondation ULB.
    Computational resources have been provided by the Consortium des Équipements de Calcul Intensif (C\'ECI), funded by the F.R.S.-FNRS under Grant No. 2.5020.11 and by the Walloon Region.
\end{acknowledgments}

\appendix
\section{Single particle spectrum}\label{app:SingleParticle}
\begin{figure}[b]
    \centering
    \includegraphics{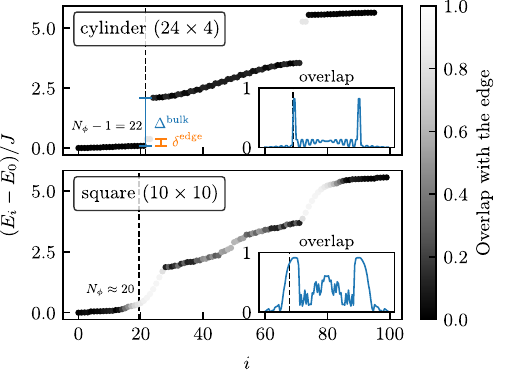}
    \caption{
        Single particle spectrum for the cylinder (top panel) and the square (bottom panel) studied here.
        The colorbar indicates the overlap of the states with the edge formed by the outermost sites of the system.
        For the cylinder we find $22$ states clearly localized in the bulk with two low-lying edge excitations at finite energy $\delta^{\rm edge}=0.28J$.
        For the square geometry is less clear, however we find approximately \mbox{$(L_x-1)(L_y-1)\alpha\approx20$} states forming a low-lying band, which we consider as the lowest Chern band here.
    }
    \label{fig:appendix:SingleParticleSpectrum}
\end{figure}
We calculated the single particle spectrum of the non-interacting Hofstadter model at flux $\alpha=\nicefrac{1}{4}$ per plaquette for the cylinder ($L_x\times L_y = 24\times 4$) and the square ($L_x=L_y=10$) studied here, see Fig.~\ref{fig:appendix:SingleParticleSpectrum}.
We also calculate the overlap of the states with the outermost sites of the system.

For the cylinder, the number of flux quanta piercing the system is exactly $N_{\phi}=(L_x-1)L_y\alpha=23$ which gives rise to $N_{\phi}-1=22$ states which have a negligible overlap with the edge and are therefore forming the bulk band.
This number of states can also be motivated by an analogy with the continuum Landau level problem on a finite cylinder.
As is typically the case for small flux $\alpha$ per plaquette, the lowest Hofstadter (bulk) band is very flat with band width $w_{\rm cylinder}\approx 0.1$.
At the same time, on a thin cylinder the energy $\delta^{\rm edge}_{\rm cylinder}\approx 0.28J$ of the two degenerate edge modes is relatively large due to finite-size effects.
The first excited bulk states occur at a bulk gap of $\Delta=2.0J$.

For the square geometry, the situation is a priori less clear since the edge of the system is much larger and bulk and edge states are lower in energy and therefore harder to identify.
While the number of flux quanta piercing the is still given by $N_{\phi}=(L-1)^2\alpha$, this number is not an integer for our choice of parameters.
Furthermore, the interplay with low lying edge states makes it hard to determine the number of bulk states in the lowest band exactly.
However, the estimate of $\lfloor N_{\phi} \rfloor=22$ states in the lowest band proved useful in earlier studies.
We consider these low-lying states as forming a bulk band of band width $w_{\rm square} \approx 0.3J$.

\section{Additional data: Square lattice}\label{app:disk}
\begin{figure}[b]
    \centering
    \includegraphics{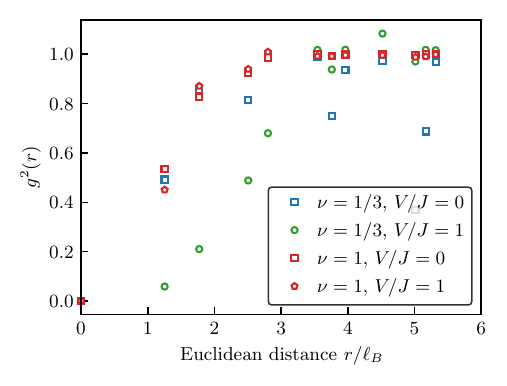}
    \caption{
        Two-particle density correlations (open boundary) as function of the Euclidean distance in units of the magnetic length \mbox{$\ell_B = \nicefrac{a}{\sqrt{2\pi \alpha}}$.}
    }
    \label{fig:NNInt:nn_corr_disk_supp}
\end{figure}
We complement the results shown in Sec.~\ref{sec:OBC} with data for all four points in the parameter space studied in this article.
$g^{(2)}(r)$ is significantly different in the conjectured FCI phase, see Fig.~\ref{fig:NNInt:nn_corr_disk_supp}.
The weak oscillations observed are remnants of the finite lattice spacing.
We conclude that the ground state is well described by the Laughlin state.
Interestingly, while we would expect the metallic correlations to differ from the ones in the CI phase, we find qualitative agreement in both states.   
\section{Additional data: Cylinders}
\subsection{Energy of the spin-polarized system}
\label{appendix:cylinder:spinpolarizedEnergetics}
\begin{figure}
    \centering
    \includegraphics{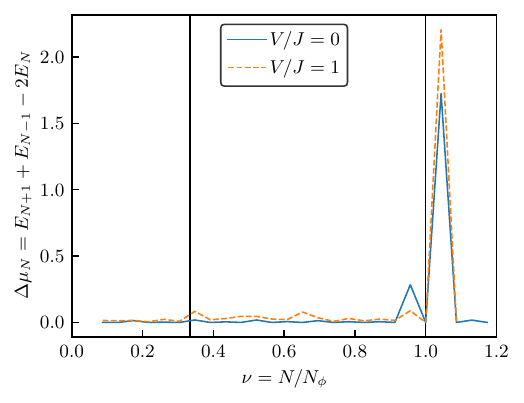}
    \caption{
        Charge gap $\Delta\mu_N$ as a proxy for the incompressibility of the ground state.
        At $\nu=1$ we find an incompressible ground state with the charge gap on the order of the single-particle gap.
        For finite nearest-neighbor interactions $\nicefrac{V}{J}=1$ we furthermore find signals of incompressibility at $\nu\approx\nicefrac{1}{3}, \nicefrac{1}{2}$, and $\nicefrac{2}{3}$, which are absent in the non-interacting case.
        While the states at $\nu\approx\nicefrac{1}{3}$ and $\nicefrac{2}{3}$ are consistent with Laughlin states, the origin and nature of the state at $\nu\approx\nicefrac{1}{2}$ is less clear.
        $L_x=24, L_y=4, \alpha=0.25$ cylinder
    }
    \label{fig:NNInt:Incompressibility}
\end{figure}

We start our discussion by considering the charge gap
\begin{equation}
    \Delta\mu_N = E_{N+1} + E_{N-1} - 2E_N
\end{equation}
as a proxy of the incompressibility of the ground state.
As expected, we find an incompressible state at $\nu=1$ with the charge gap of the order of the single particle band gap $\Delta$, see Fig.~\ref{fig:NNInt:Incompressibility}.
In the interacting case, we furthermore find signs of incompressibility at $\nu\approx\nicefrac{1}{3}, \nicefrac{1}{2}$, and $\nicefrac{2}{3}$.
%
% While the states at $\nu\approx\nicefrac{1}{3}$ and $\nicefrac{2}{3}$ are consistent with Laughlin states, the origin and nature of the state at $\nu\approx\nicefrac{1}{2}$ is less clear.
% %
% \textit{\color{lightgray}Felix' thoughts: Not sure how much we want to say here about $\nu=\nicefrac{1}{2}$. We could also just leave it as a funny feature of the model. Maybe finite-size studies can remove this signal anyway?}

\begin{figure}[t!]
    \centering
    \includegraphics{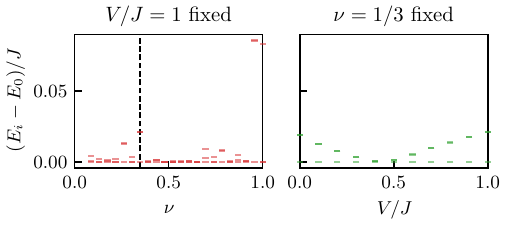}
    \caption{
        Low-lying spectrum for cuts along characteristic lines in parameter space.
        Most prominently, we observe an enhancement of the many-body gap at $\nu = \nicefrac{1}{3}$ (dashed line) with a unique ground state.
        %
        % We observe characteristic degeneracies in the spectrum, especially around $\nu=\nicefrac{1}{3}$.
        % %
        % In particular, we observe an opening of the many-body gap above the two-fold degenerate ground state at $\nu=\nicefrac{1}{3}$.
        %
        For $\nu=\nicefrac{1}{3}$, the gap grows monotonously with $V$ for $\nicefrac{V}{J} \gtrsim 0.5$.
        $L_x=24, L_y=4, \alpha=0.25$ cylinder
    }
    \label{fig:NNInt:GS_Degeneracy}
\end{figure}
To investigate the energetics of the topological candidate state further, we calculate the variational energy of the first three low-lying excited states, see Fig.~\ref{fig:NNInt:GS_Degeneracy}.
We find that at $\nu=\nicefrac{1}{3}$ the many-body gap closes around $\nicefrac{V}{J}\approx 0.5$, in agreement with the observables discussed in the main text.
At the same time, for fixed interactions $\nicefrac{V}{J}=1$ we find an enhancement of the many-body gap at $\nu\approx \nicefrac{1}{3}, \nicefrac{1}{2}, \nicefrac{2}{3}$ and $1$, in agreement with our incompressibility scans.
We take this as further evidence for the $\nicefrac{1}{3}$-Laughlin-like nature of the strongly interacting ground state at $\nu=\nicefrac{1}{3}$.

\begin{figure}[b]
    \centering
    \includegraphics{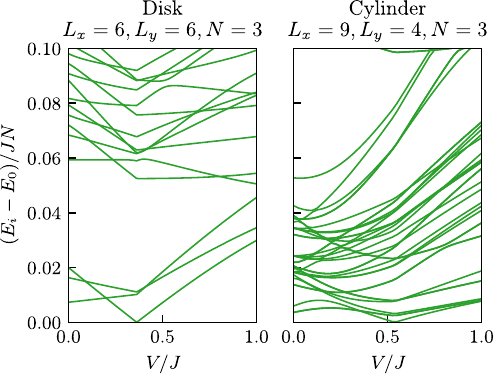}
    \caption{
        Many-body spectrum obtained by exact diagonalization of a few particles on a open boundary square lattice geometry (left) and a cylinder (right).
        In both cases, we find a gap closing around $\nicefrac{V}{J}\approx 0.5$, in agreement with our DMRG simulations.
        For the cylinder we furthermore find a very dense low-energy spectrum below this critical point which might evolve into a gapless, metallic spectrum in the thermodynamic limit.
    }
    \label{fig:NNInt:GapOpening_ED}
\end{figure}

The opening of a many-body gap around $\nicefrac{V}{J}\approx 0.5$ is further confirmed by exact diagonalization of small systems, see Fig.~\ref{fig:NNInt:GapOpening_ED}.
Furthermore, the spectrum on a cylinder below the critical point hints towards the formation of a gapless, metallic spectrum in the thermodynamic limit.
In contrast, in the strongly interacting regime there are clearly separated low-lying excited states.

\subsection{Spin-spin correlations in the Hubbard model ($\nicefrac{V}{J}=0$)}
\label{appendix:cylinder:pureHubbard:spinSector}
\begin{figure}[b]
    \centering
    \includegraphics{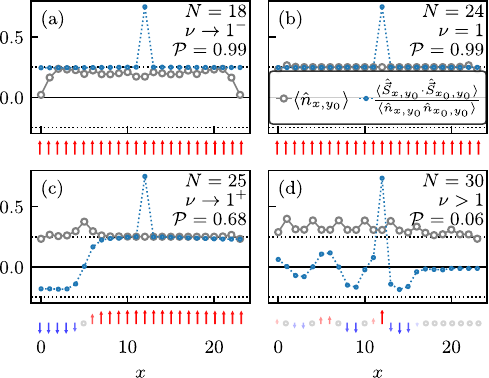}
    \caption{
        Normalized spin-spin correlations for \mbox{$\nicefrac{U}{J}=8$}, $\nicefrac{V}{J}=0$ and different numbers $N$ of particles.
        Below each panel, the corresponding spin configuration is sketched.
        We observe that the long-range ferromagnetic correlations at $\nu\lesssim1$ (a and b) evolve into non-trivial, skyrmion-like spin textures carried by particle excitations at $\nu \to 1^+$ (c).
        Upon adding even more particles, the correlations exhibit short range spin anti-alignment (d).
        Note that for $\mathbf{i}=\mathbf{j}$ we expect $\braket{\vec{S}_{\mathbf{i}}\cdot\vec{S}_{\mathbf{i}}} = s(s+1) = \nicefrac{3}{4}$ based on the spin-1/2 of the constituent particles.
    }
    \label{fig:appendix:cylinder:SpinCorrelations_V0}
\end{figure}
We complement our studies of the spin-spin correlations close to $\nu=1$ with simulations of the vanilla Hubbard model, i.e. in the absence of nearest-neighbor repulsion.
For $\nu\geq 1$ we find qualitatively the same behavior as for the extended model, see Fig.~\ref{fig:appendix:cylinder:SpinCorrelations_V0}(b-d) compared to Fig.~\ref{fig:cylinder:SpinCorrelations_V1}(f-h).
However, for $\nu < 1$ we do not find any evidence of hole-skyrmions, see Fig.~\ref{fig:appendix:cylinder:SpinCorrelations_V0}(a), but instead find long-range ferromagnetic correlations.
This is also in agreement with the fully spin-polarized ground state discussed in the main text and with earlier studies~\cite{Palm2023,Ding2024}.

\end{document}